\documentclass[twocolumn,amsmath,amssymb,floatfix,preprintnumbers,prl,showpacs,lengthcheck,superscriptaddress]{revtex4-1}
\usepackage{graphicx}
\usepackage{amsmath}
\usepackage{amssymb}
\usepackage{hyperref}
\usepackage[capitalise]{cleveref}

\newcommand{\beq}{\begin{equation}}
\newcommand{\eeq}{\end{equation}}
\newcommand{\beqs}{\begin{eqnarray}}
\newcommand{\eeqs}{\end{eqnarray}}

\newcommand{\pbp}[0]{\ensuremath{\langle \overline{\psi} \psi \rangle}}

\newcommand{\chidof}{\ensuremath{\mbox{$\chi^2/\text{d.o.f.}$}}}

\newcommand{\mres}{m_{\rm res}}

\begin{document}

\preprint{FERMILAB-PUB-12-111-T}
\preprint{LLNL-JRNL-548639}
\preprint{NSF-KITP-12-069}

\title{Approaching Conformality with Ten Flavors}

\author{T.~Appelquist}
\affiliation{Department of Physics, Sloane Laboratory, Yale University,
             New Haven, Connecticut 06520, USA}
\affiliation{Kavli Institute for Theoretical Physics, University of California, Santa Barbara, California 93106, USA}
\author{R.~C.~Brower}
\affiliation{Department of Physics, Boston University,
	Boston, Massachusetts 02215, USA}
\affiliation{Kavli Institute for Theoretical Physics, University of California, Santa Barbara, California 93106, USA}
\author{M.~I.~Buchoff}
\affiliation{Physical Sciences Directorate, Lawrence Livermore National Laboratory,
	Livermore, California 94550, USA}
\affiliation{Kavli Institute for Theoretical Physics, University of California, Santa Barbara, California 93106, USA}
\author{M.~Cheng}
\affiliation{Department of Physics, Boston University,
	Boston, Massachusetts 02215, USA}
\author{S.~D.~Cohen}
\affiliation{Department of Physics, University of Washington, Box 351560,
               Seattle, WA 98195, USA}
\author{G.~T.~Fleming}
\affiliation{Department of Physics, Sloane Laboratory, Yale University,
             New Haven, Connecticut 06520, USA}
\affiliation{Kavli Institute for Theoretical Physics, University of California, Santa Barbara, California 93106, USA}
\author{J.~Kiskis}
\affiliation{Department of Physics, University of California,
	Davis, California 95616, USA}
\author{M.~F.~Lin}
\affiliation{Department of Physics, Sloane Laboratory, Yale University,
             New Haven, Connecticut 06520, USA}
\author{H.~Na}
\affiliation{Argonne Leadership Computing Facility,
	Argonne, Illinois 60439, USA}          
\author{E.~T.~Neil}
\affiliation{Theoretical Physics Department, Fermi National Accelerator Laboratory,
             Batavia, IL 60510, USA}
\affiliation{Kavli Institute for Theoretical Physics, University of California, Santa Barbara, California 93106, USA}
\author{J.~C.~Osborn}
\affiliation{Argonne Leadership Computing Facility,
	Argonne, Illinois 60439, USA}
\author{C.~Rebbi}
\affiliation{Department of Physics, Boston University,
	Boston, Massachusetts 02215, USA}
\author{D.~Schaich}
\affiliation{Department of Physics,
University of Colorado, Boulder, CO 80309, USA}
\author{C.~Schroeder}
\affiliation{Physical Sciences Directorate, Lawrence Livermore National Laboratory,
	Livermore, California 94550, USA}
\author{G.~Voronov}
\affiliation{Department of Physics, Sloane Laboratory, Yale University,
             New Haven, Connecticut 06520, USA}
\author{P.~Vranas}
\affiliation{Physical Sciences Directorate, Lawrence Livermore National Laboratory,
	Livermore, California 94550, USA}
\affiliation{Kavli Institute for Theoretical Physics, University of California, Santa Barbara, California 93106, USA}
\collaboration{Lattice Strong Dynamics (LSD) Collaboration}
\noaffiliation

\begin{abstract}
We present first results for lattice simulations, on a single volume, of the low-lying spectrum of an SU$(3)$ Yang-Mills gauge theory with $N_f = 10$ light fermions in the fundamental representation.  Fits to the fermion mass dependence of various observables are found to be globally consistent with the hypothesis that this theory is within or just outside the strongly-coupled edge of the conformal window, with mass anomalous dimension $\gamma^\star$ consistent with 1 over the range of scales simulated.  We stress that we cannot rule out the possibility of spontaneous chiral-symmetry breaking at scales well below our infrared cutoff.  We discuss important systematic effects, including finite-volume corrections, and consider directions for future improvement.
\end{abstract}


\maketitle

\paragraph{\textbf{Introduction}}

In three recent papers \cite{Appelquist:2009ka, Appelquist:2010xv, Appelquist:2012sm}, we studied the properties of an SU$(3)$ gauge theory with $N_f$ massless Dirac fermions in the fundamental representation as $N_f$ increases from $2$ to $6$. We noted that the $N_f = 2$ simulations are in good agreement with measured QCD values, and that the $N_f = 6$ results indicate substantial enhancement of the chiral condensate \cite{Appelquist:2009ka}. We also observed  that the spectrum becomes more parity doubled and the $S$ parameter per electroweak doublet decreases as $N_f$ increases from $2$ to $6$ \cite{Appelquist:2010xv}. Most recently, we examined $\pi - \pi$ scattering in these theories \cite{Appelquist:2012sm}, noting that the low-energy scattering length decreases when $N_f$ increases from $2$ to $6$.

The $N_f = 6$ theory is thought to exhibit these trends because $N_f$ has increased toward a critical value $N_f^c$ at which the theory transitions from confinement and chiral symmetry breaking to infrared conformal behavior. The precise value of $N_f^c$ remains uncertain, but all recent evidence points to confinement and chiral-symmetry breaking for an $SU(3)$ theory with $N_f = 8$ \cite{Appelquist:2007hu,Appelquist:2009ty,Hasenfratz:2010fi,Aoki:2012ep}, while most recent analyses indicate that an $SU(3)$ theory with $N_f = 12$ is conformal in the infrared \cite{Appelquist:2007hu,Appelquist:2009ty,Aoyama:2011ry,Appelquist:2011dp,DeGrand:2011cu,Aoki:2012kr,Deuzeman:2012pv,Hasenfratz:2011xn} (although some studies have concluded that it is spontaneously broken \cite{Fodor:2011tu,Jin:2012dw}).

Here, we summarize results of numerical lattice simulations for an $SU(3)$ gauge theory with $10$ light Dirac fermions in the fundamental representation. This theory was recently studied by Yamada et al \cite{Yamada:2010wd, Hayakawa:2010yn,10f-update}, who computed the running gauge coupling in the Schr\"{o}dinger-functional scheme, finding evidence that the theory is conformal in the infrared, dominated by a relatively strong fixed point and with a large mass anomalous dimension.

We examine the particle spectrum of the $N_f = 10$ theory, including the masses and decay constants of the lowest-lying pseudoscalar, vector and axial-vector states, and the masses of the lowest-lying nucleon state and its parity partner. We also consider the chiral condensate.  We compute these quantities on the lattice using a finite fermion mass $m$, and analyze the behavior as $m$ is extrapolated toward zero.  

Motivated by the running coupling results, we focus first on the infrared conformal hypothesis. We argue that the masses, decay constants, and chiral condensate scale with $m$ over a range of $m$ values in a manner consistent with mass-deformed conformal perturbation theory \cite{Miransky:1998dh,Luty:2008vs,DelDebbio:2010jy,DelDebbio:2010ze,Appelquist:2011dp}, indicating that the theory is either inside or just below the edge of the conformal window.  We cannot rule out the possibility that the theory will exhibit an intrinsic confinement scale and spontaneous chiral-symmetry breaking scale at lower values of $m$.  As $m$ is decreased, finite-volume effects become more important, and these effects must be considered carefully, particularly since our current results are restricted to a single volume.

\paragraph{\textbf{Simulation details}}

Simulations are performed using domain-wall fermions and the Iwasaki improved gauge action \cite{Allton:2008pn}.  The domain wall formulation suppresses the chiral symmetry breaking associated with fermion discretization, and preserves flavor symmetry at finite lattice spacing. Dimensionful quantities are given in lattice units, with implicit dependence on the lattice spacing $a$.  The lattice volume is set to $32^3 \times 64$, with the length of the fifth dimension $L_s = 16$ and the domain wall height $m_0 = 1.8$. We choose $\beta \equiv 6/g_0^2 = 1.95$, which lies on the weak-coupling side of a bulk phase transition, and leads to a vector-meson mass $0.24 \leq M_V \leq 0.34$ for the range of fermion masses $m$ used in our analysis.  This is similar to our previous studies at $N_f = 2$ and $N_f = 6$~\cite{Appelquist:2009ka, Appelquist:2010xv}.  Each ensemble generated contains approximately 1200 gauge configurations.

 Input fermion masses $m_f = 0.01$ to $0.03$ are included in the simulations.  At finite lattice spacing, even with $m_f = 0$, the chiral symmetry is not exact, with the violation captured in a residual mass $\mres$. The total fermion mass $m$ is then  $ m \equiv m_f + \mres$. For our simulations, $\mres \approx 0.0017$, so that $\mres \ll m_f$ for all values of $m_f$.

To study the effects of thermalization and fixed topological charge, we have generated gauge field configurations from both disordered and ordered starts for most ensembles.  We note that an $m_f = 0.005$ ensemble was also generated, but did not show signs of adequate thermalization over the number of configurations generated, and we will not discuss those results further here.

\paragraph{\textbf{Combination of ordered/disordered data} \label{sec:combine}}

The topological charge $Q$ is observed to evolve very slowly on all ensembles, so that disordered starts generally remain in a sector with large net topological charge, while ordered starts are essentially stuck in a sector with zero net topological charge.  On the subset of our ensembles where extrapolation of $Q$ for disordered starts is possible, we find that these effects are able to explain most of the observed discrepancy between observable values on the two ensembles.  In a future study currently in progress, we will include these topological-charge corrections explicitly by measuring the topological susceptibility on all ensembles.

Since neither of our evolutions at a given mass point has sufficient topological tunneling, we combine the results using the difference to estimate a systematic error.  For an observable $\mathcal{O}$, determined on a pair of ensembles with the same physical parameters but with different initial states (ordered and disordered), we assume that the samples of $\mathcal{O}$ are sufficiently large within a given ensemble that the central limit theorem applies.  Each distribution is then Gaussian, with mean and standard error $(\mu_1, \sigma_1), (\mu_2, \sigma_2)$.  There remains an unknown bias of the mean computed within a given topological sector with respect to the true mean $\hat{\mu}$, resulting from a properly-weighted distribution over all sectors.

We expect that the true mean lies somewhere between $\mu_1$ and $\mu_2$.  To obtain a conservative estimate of the true distribution, we take a uniform distribution of width $\delta$ and center $\bar{\mu} = (\mu_1 + \mu_2) / 2$ to describe our knowledge of the bias-corrected mean.  Convolving this uniform distribution with the Gaussian sample distributions, the combined mean is given by $\mu_c = \bar{\mu}$, while the variance is equal to $\sigma_c^2 = \frac{1}{2}(\sigma_1^2 + \sigma_2^2) + \frac{1}{3} \delta^2$.  We determine the width $\delta$ by assuming that the width as a fraction of the mean is constant for a given observable as a function of light fermion mass, taking the maximum fractional difference over the values observed as the best estimate.  The resulting $\delta$ values range from $5$-$15$\% for most observables, so that the $\delta$ contribution to the variance $\sigma_c^2$ is always significant.

\begin{figure}
\includegraphics[width=85mm]{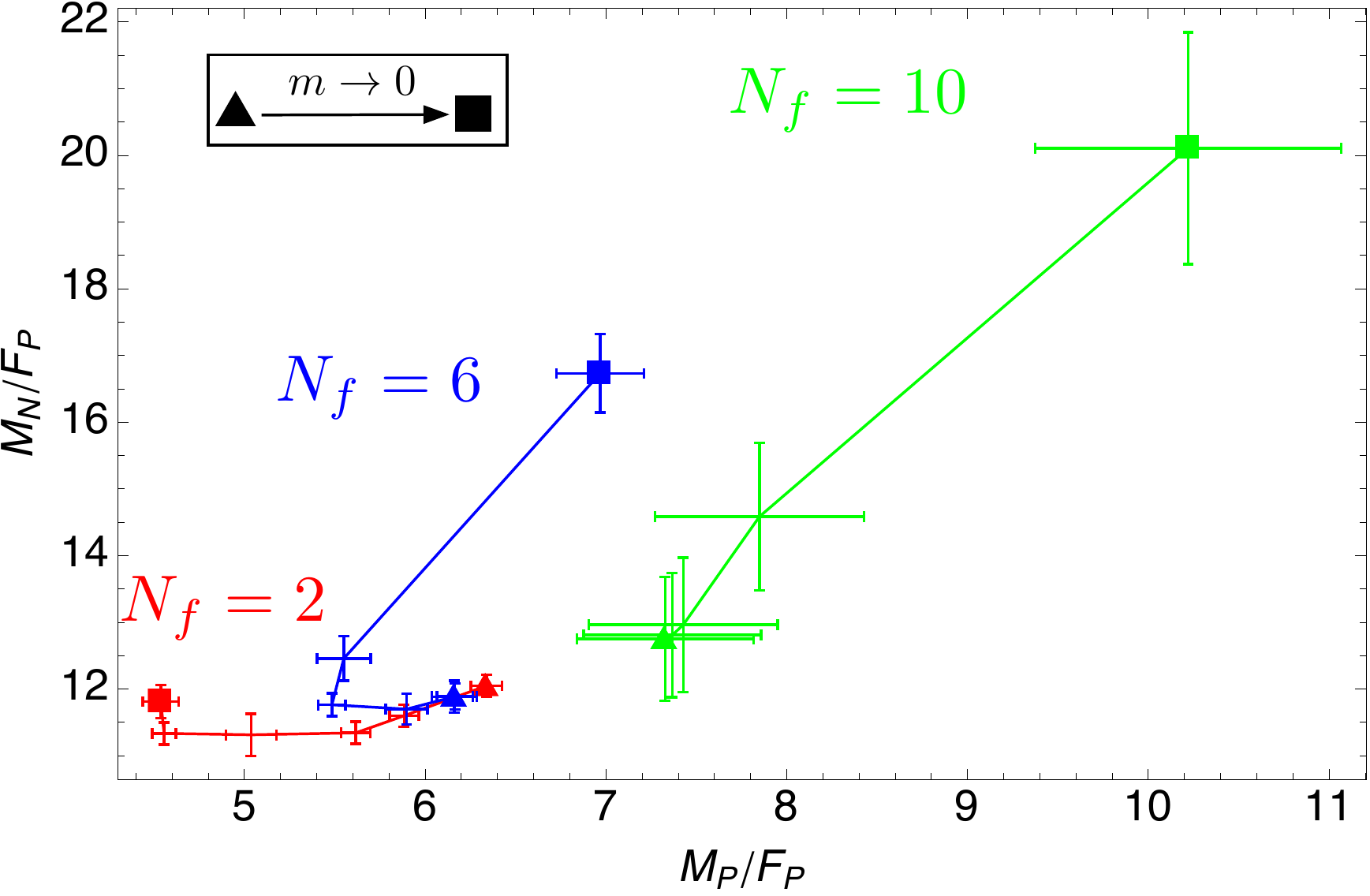}
\caption{Edinburgh-style plot comparing ratios of physical observables for simulations of the $N_f = 2$ (red), $N_f = 6$ (blue), and $N_f = 10$ (green) theories.  The points are joined together in decreasing order in the fermion mass, with the heaviest and lightest masses denoted by a triangle and box respectively.  The combinations of observables plotted are $M_N / F_P$ vs. $M_P / F_P$, chosen to clearly illustrate finite-volume corrections which act with opposite sign on bound-state masses and $F_P$. \label{fig:edinburgh}}
\end{figure}

\paragraph{\textbf{Finite-volume effects}}

Our simulations so far have used a single lattice spatial volume, $L/a = 32$. Finite-volume effects can in principle be incorporated through a controlled expansion in some function of $ML$, with $M$ the relevant mass scale for a particular framework as in Ref. \cite{Appelquist:2011dp}.  Here we will discuss finite-volume effects qualitatively, and argue that such effects should be relatively small on a subset of our results.

An estimate of the significance of finite-volume corrections can be made by comparing ratios of physical observables.  Edinburgh-style plots, which were used frequently as a diagnostic tool in the early literature on lattice QCD \cite{Bowler:1985hr}, can be particularly useful for theories other than QCD \cite{Deuzeman:2012pv}.  In \cref{fig:edinburgh}, we compare the ratios of observables $M_N / F_P$ and $M_P / F_P$.  In order to better illustrate the expected behavior, we include our own spectrum measurements for $N_f = 2$ and $6$ \cite{Appelquist:2009ka,Appelquist:2010xv} for comparison.

The combinations $M_N / F_P$ and $M_P / F_P$ are chosen to clearly show finite-volume corrections.  Decreasing the lattice volume tends to increase $M_N$ and $M_P$ while simultaneously decreasing $F_P$ relative to their values in the infinite-volume limit, driving the points up and to the right in the plot as $m \rightarrow 0$.  On the other hand, the infinite-volume scaling behavior for a chirally broken theory has the points moving to the left as $m \rightarrow 0$, as the pion mass scales to zero.  For the $N_f = 6$ case, the latter trend changes into the former around $m_f = 0.010$.

For the $N_f = 10$ data combined using the method described above, no movement is seen within errors for $m_f \geq 0.020$, consistent with the possibility that $N_f = 10$ can be described by a mass-deformed conformal expansion.  In this expansion, all scales vanish as a common power of $m$ at leading order, so that ratios of observables remain fixed as $m$ is varied.  The $m_f = 0.015$ and $m_f = 0.010$ points show small and large displacements, respectively, which may indicate  finite-volume effects becoming large at $m_f = 0.010$.  The fits described in the next section, for $m_f \geq 0.010$, $ \geq 0.015$ and $ \geq 0.020$, are consistent with these conclusions. 

The lack of scaling with $m$ of the points on this plot is not itself sufficient evidence that the mass-deformed conformal framework is an adequate description of the $N_f = 10$ results.  For a theory in a heavy-quark regime with all observables showing linear dependence on $m$, the ratios shown are also expected to approach a constant value for large $m$.  However, the consistency of this plot with expectations from the conformal hypothesis motivates further study within that framework.  It is also clear from \cref{fig:edinburgh} that chiral perturbation theory cannot provide an accurate description of our $N_f = 10$ results: there are no signs of the expected decrease of $M_P$ with respect to $F_P$, and the numerical size of $M_P / F_P$ is too large for the chiral expansion to be convergent.

\paragraph{\textbf{Infrared conformal hypothesis}}

If the $N_f = 10$ theory is conformal in the infrared limit, then the chiral symmetry is broken only by the explicit fermion mass, and the expected mass dependence of the spectrum is determined by the emergent conformal symmetry. This possibility is supported by evidence from running-coupling studies \cite{Hayakawa:2010yn,Yamada:2010wd} that the gauge coupling $g^{2}(\mu)$ evolves slowly at long distances.  Therefore, we fit the spectrum using the infrared conformal hypothesis, assuming that $g^{2}(\mu)$ remains at its fixed-point value $g^{{\star}2}$, and that the data can therefore be described by mass-deformed conformal field theory \cite{Miransky:1998dh, Luty:2008vs, DelDebbio:2010jy, DelDebbio:2010ze, Appelquist:2011dp} with mass anomalous dimension $\gamma^{\star}$.

When an explicit fermion mass $m \equiv m(\Lambda) \ll \Lambda$ is introduced, the running mass for scales below $\Lambda$ is given by  $m(\mu) = m (\Lambda / \mu)^{\gamma^\star}$.  For some energy scale $M \ll \Lambda \sim 1/a$, the running mass thus satisfies the equality $m(M) = M$, so that at scales small compared to $M$ the fermions decouple from the theory, leaving an effective pure-gauge theory which confines as the gauge coupling flows away from the fixed-point value $g^{\star 2}$.  So long as $g^{\star 2}$ is reasonably strong, the induced confinement scale will be of order $M$.  The mass of all fermion bound states is then given by \cite{Appelquist:2011dp}
\beq
M_X = C_{X}~ m^{[1 /( 1+ \gamma^{\star})]} + D_{X}~  m, \label{eq:confM}
\eeq
where we have included a small correction term. With the masses expressed in units of the cutoff $\Lambda$,  $C_X$ and $D_X$ are dimensionless coefficients. Since the explicit breaking of chiral symmetry is of order $M$, there is no approximate chiral symmetry to be broken spontaneously. Thus this scaling law applies as well to the pseudoscalar mass.

\begin{figure}
\includegraphics[width=85mm]{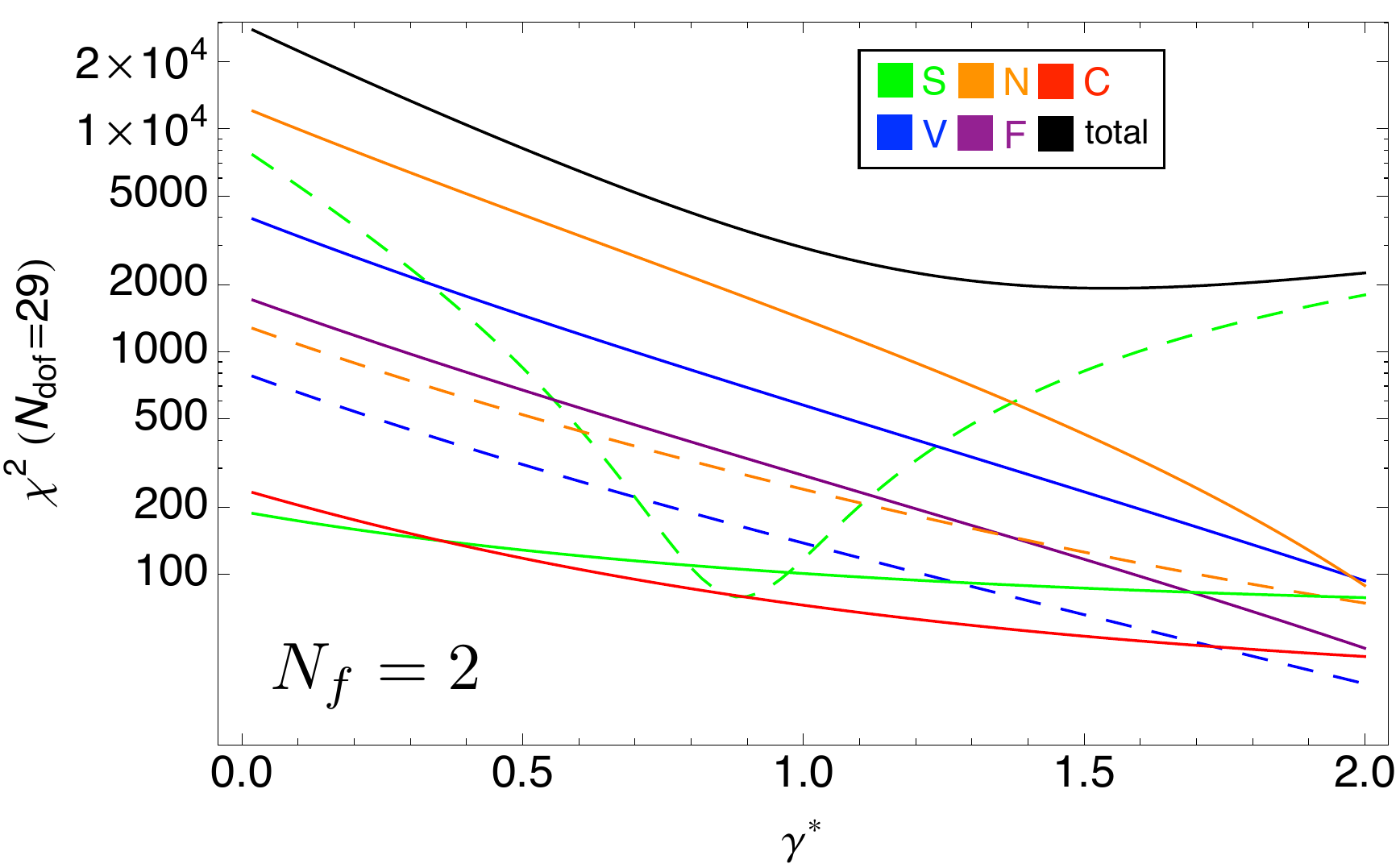}
\includegraphics[width=85mm]{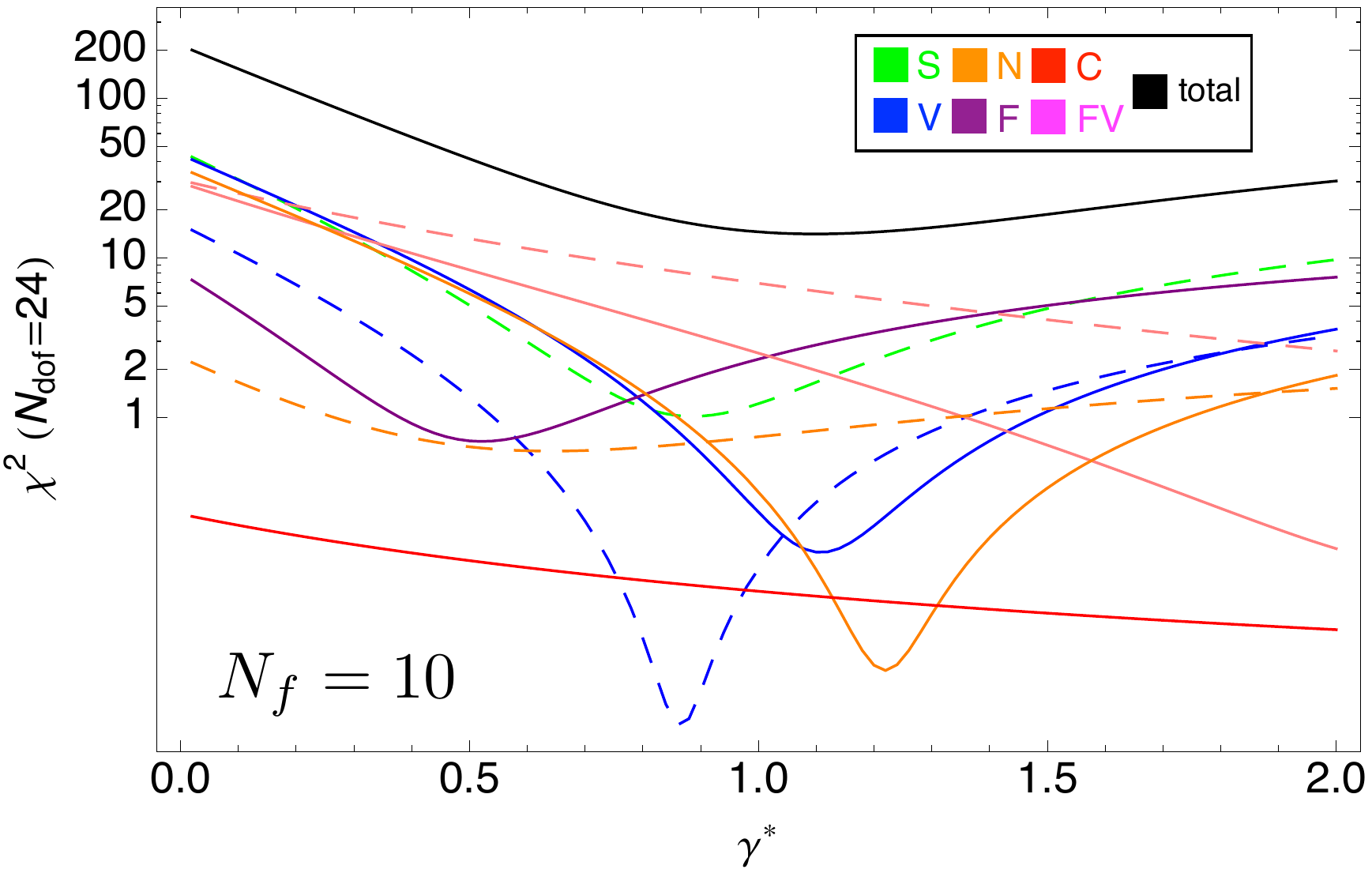}
\caption{$\chi^2$ scans as a function of $\gamma^\star$ for our $N_f = 2$ results (top) and $N_f = 10$ results in the range $m_f \geq 0.015$ (bottom).  Contours shown are, from bottom to top at $\gamma^\star = 0$ and $N_f = 10$: $\pbp$ (red), $M_{N*}$ (orange, dashed), $F_P$ (purple), $M_A$ (blue, dashed),  $F_V$ (pink), $F_A$ (pink, dashed), $M_N$ (orange), $M_V$ (blue), $M_P$ (green, dashed), and total $\chi^2$ (black).  At $N_f = 2$, the pseudoscalar mass shows the expected scaling behavior $M_P^2 \sim m$, which appears as a minimum at $\gamma^\star \approx 1$ in this analysis.\label{fig:chi2scan}}
\end{figure}

The pseudoscalar, vector, and axial-vector decay constants as we define them are expected to scale in the same way as bound-state masses \cite{DelDebbio:2010ze}.  The chiral condensate has a more complicated dependence on the fermion mass \cite{Appelquist:2011dp}:
\begin{align}
\langle{\bar\psi} \psi \rangle &= A_{C}m  +  B_{C} m^{[(3 - \gamma^{\star}) /( 1+ \gamma^{\star})]}  \nonumber \\
&+ C_{C} m^{[3 /( 1+ \gamma^{\star})]}  + D_{C} m^3. \label{eq:confC}
\end{align}
 The above expressions vanish as $m \rightarrow 0$ with the scaling determined by a single parameter $\gamma^\star$, a behavior qualitatively different from that of a theory with spontaneous chiral symmetry breaking. 
 
Results for global fits to the combined simulation data for the ranges $m_f \geq 0.010$, $m_f \geq 0.015$ and $m_f \geq 0.020$ are shown in \cref{tab:fitconf}.  Since we have a relatively small number of mass points to work with, we here consider only fits with the  $D$-terms set to zero.  As anticipated, the fit quality is reasonably good for the restrictions $m_f \geq 0.015$ and $m_f \geq 0.020$.  Including the $m_f = 0.010$ data changes the fit parameters significantly, matching our expectation that finite-volume corrections become large for these points.  The other two fits indicate a large anomalous dimension $\gamma^\star \gtrsim 0.8$ and are consistent with $\gamma^\star = 1$, the value anticipated for a theory with $N_f$ at the edge of the conformal window \cite{Holdom:1984sk,Yamawaki:1985zg,Appelquist:1986an}. 

To better understand our fit results, we show in \cref{fig:chi2scan} scans over $\chi^2$ as a function of $\gamma^\star$, broken up for each individual observable included in the $m_f \geq 0.015$ fit.  Several of the observables show an individual minimum in $\chi^2$ compatible with the global best-fit value $\gamma^\star \approx 1.10$.  The chiral condensate (shown in red) has no clear minimum, but it contributes very little to overall $\chi^2$, so we omit it from the global fits in \cref{tab:fitconf}.  The global fit with the condensate included is not significantly different, but exhibits very strong correlations between the parameters $\gamma^\star$, $A_C$ and $B_C$; this behavior is expected for $\gamma^\star$ near 1, for which the $A_C$ and $B_C$ terms in \cref{eq:confC} are nearly degenerate.

As is evident from \cref{fig:chi2scan}, the distribution of $\chidof$ as a function of $\gamma^\star$ is not symmetric about the minimum.  We estimate a two-sided 68\% (95\%) confidence interval on $\gamma^\star$ directly, varying by $\Delta \chi^2 = 1$ ($\Delta \chi^2 = 4$) about the minimum of the $\chi^2$ contour shown in \cref{fig:chi2scan}.  Results for each mass range are shown in \cref{tab:fitconf}.  In all cases we find $\gamma^\star \gtrsim 0.8$ at two sigma.

A similar plot using our $N_f = 2$ results is shown for comparison.  As expected the $N_f = 2$ theory shows generally very poor power-law fits for any $\gamma^\star < 2$, with the exception of the pseudoscalar mass (green, dashed), which scales as $M_P^2 \sim m$ in accordance with chiral perturbation theory.

\begin{table}
\begin{tabular}{|c|c|c|c|}
\hline
\textbf{Obs.} & $m_f \geq 0.010$ & $m_f \geq 0.015$ &$m_f \geq 0.020$ \\
\hline\hline
$\gamma^\star$&\textit{1.69(16)}&1.10(17)&1.35(47)\\
$[ 68\%$ CI$]$ &\textit{[1.54,1.86]}&[0.95,1.27] &[1.06,1.73]\\
$[ 95\%$ CI$]$ &\textit{[1.40,2.06]}&[0.82,1.46] &[0.83,2.27]\\
\hline\hline
$C_P$&\textit{0.98(9)}&1.44(21)&1.21(37)\\
$C_V$&\textit{1.17(10)}&1.70(25)&1.42(44)\\
$C_A$&\textit{1.43(13)}&2.14(32)&1.79(56)\\
$C_N$&\textit{1.75(16)}&2.53(37)&2.10(65)\\
$C_{N^\star}$&\textit{2.23(25)}&3.35(55)&2.87(92)\\
$C_{FP}$&\textit{0.121(12)}&0.190(28)&0.164(51)\\
$C_{FV}$&\textit{0.165(15)}&0.238(35)&0.195(60)\\
$C_{FA}$&\textit{0.136(13)}&0.192(28)&0.154(48)\\
\hline\hline
$\chidof$&\textit{69/31}&14/23&3.1/15\\
\hline
\end{tabular}
\caption{Global fit results for the conformal hypothesis of \cref{eq:confC,eq:confM}, based on combined ordered/disordered data as described in the text.  The labels $P, V, A, N, N^\star$ correspond to the pseudoscalar, vector, axial-vector, nucleon and nucleon-prime, respectively.  Decay constants for channel $X$ are denoted by $FX$.  Errors shown on all quantities are purely statistical, and ignore correlations between observables.  For $\gamma^\star$, two-sided 68\% and 95\% confidence intervals are also shown.  The $m_f \geq 0.010$ fit (left column) has significantly worse $\chidof$, possibly due to the presence of finite-volume effects.\label{tab:fitconf}}
\end{table}

\paragraph{\textbf{Chirally Broken Hypothesis}}

Despite the quality of fits obtained under the infrared-conformal hypothesis, it remains possible that the $N_f = 10$ theory is chirally broken. A rigorous test of this possibility would involve chiral perturbation theory to extrapolate to $m = 0$. But as discussed in the context of \cref{fig:edinburgh}, we do not expect this expansion to be convergent for $m_f \geq 0.015$.  We have nevertheless attempted to fit our $N_f = 10$ results using NLO chiral perturbation theory, as done previously for $N_f = 2$ \cite{Appelquist:2009ka} and $N_f = 6$ \cite{Neil:2010sc}, finding (at $N_f = 10$) generally large values of $\chidof$ and best-fit values pointing to a poorly convergent expansion.  We omit the details of these fits here, but will present them in a future work.

\begin{figure}[t]
\includegraphics[width=90mm]{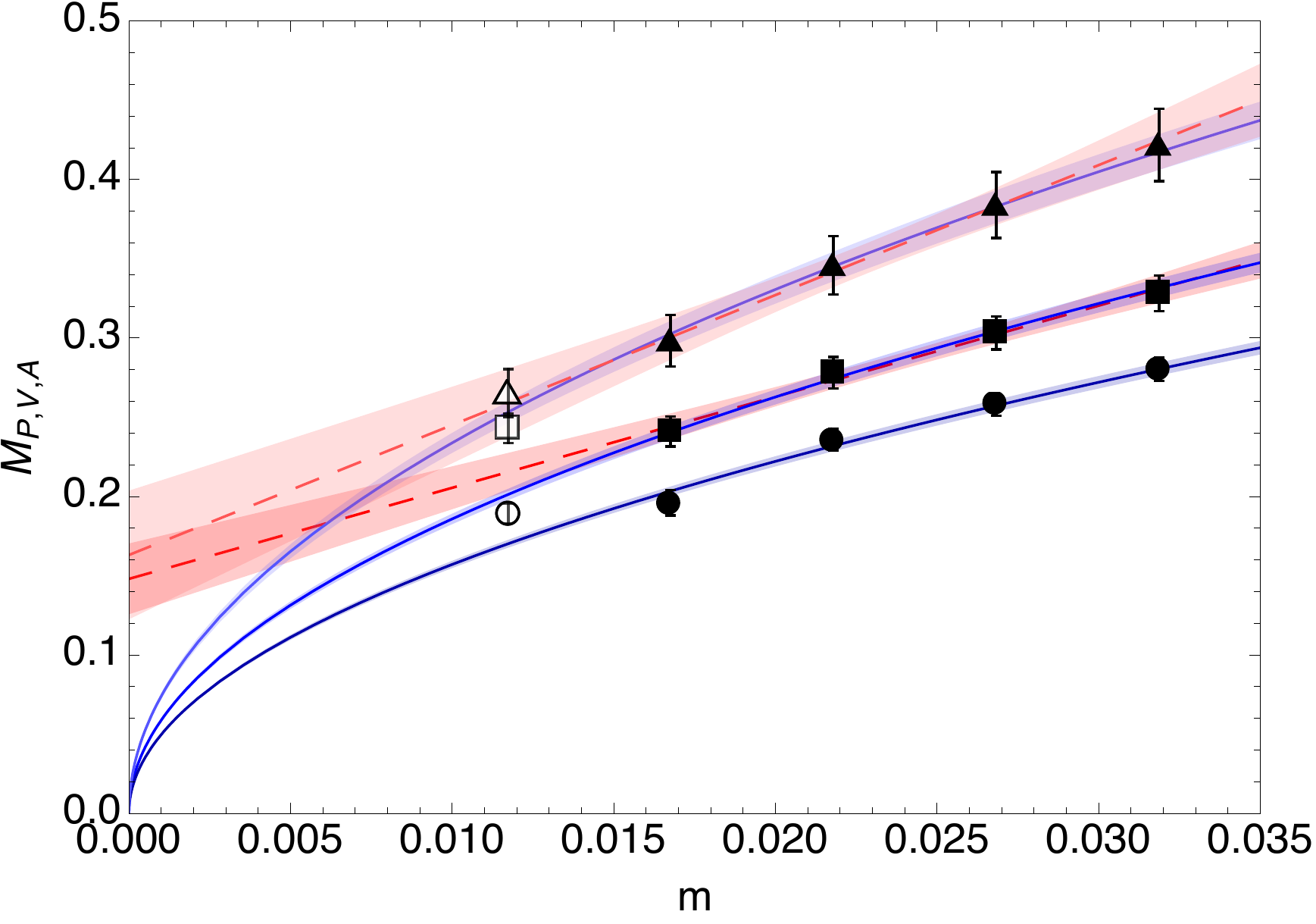}
\caption{Simulation results for the pseudoscalar mass (circles), vector mass (squares), and axial-vector mass (triangles).  Error bars on the points are estimated using the combination method described in the text.  Two fit types are compared: linear $M_V \sim a_V + b_V m$ (red), and power law $M_V \sim m^{1/2}$ (blue).    The power-law fits correspond to a mass-deformed conformal theory with $\gamma^\star = 1$.  Only observable values with $m_f \geq 0.015$ (filled symbols) are used in the fits.  \label{fig:fitcmp}}
\end{figure}

An alternative, crude approach is to use the extrapolation formula $M_P \sim b_{P} m^{1/2}$ for the pseudoscalar mass, and the linear expression $M_X \sim a_X + b_X m$ for the other masses and decay constants. In \cref{fig:fitcmp}, we compare fits of this type for the vector and axial-vector masses to mass-deformed conformal fits with fixed $\gamma^\star = 1$, a value within the errors of our global conformal fit.  Within the range of fermion masses considered, we cannot clearly distinguish this simple linear dependence from the power-law behavior of the mass-deformed conformal fits based on our current results.  We also show fit results for the pseudoscalar mass; under either the conformal or chirally broken hypothesis, this state scales as $M_P \sim m^{1/2}$, so only a single fit is shown.

\paragraph{\textbf{Discussion}}

We have presented here the first non-perturbative calculation of the spectrum for an SU$(3)$ gauge theory with $N_f = 10$ light fermions in the fundamental representation.  Previous studies of the running coupling in this theory have indicated the presence of a strongly-coupled infrared fixed point \cite{Yamada:2010wd, Hayakawa:2010yn} with mass anomalous dimension  $\gamma^\star \sim 1$ \cite{10f-update}.  Our simulation results are found to be consistent with infrared conformality, up to large finite-volume corrections which appear at our lightest fermion masses $m_f \leq 0.010$.  A global fit using the framework of mass-deformed conformal perturbation theory \cite{Miransky:1998dh, Luty:2008vs, DelDebbio:2010jy, DelDebbio:2010ze, Appelquist:2011dp} yields a best-fit value $\gamma^\star = 1.10(17)$ (with only statistical error shown). 

The mass-deformed conformal fit carried out here is rather simple, for example setting $g(\mu) \approx g^\star$ and thus ignoring weakening of the gauge coupling at energies near the UV cutoff. Our numerical estimates for $\gamma^\star$ should therefore be regarded with caution.  More importantly, we cannot rule out the possibility that at some energy below the induced confinement
scale $M$ corresponding to $m_f \simeq 0.015$, intrinsic confinement and spontaneous chiral symmetry breaking would set in.  The mass-dependence of all observables would then be expected to switch continuously from power-law scaling to the forms expected in chiral perturbation theory.

In future work, simulations at additional fermion masses and on additional volumes will be crucial. If the $N_f = 10$ theory is indeed conformal or near-conformal in the infrared, all bound-state masses are expected to follow universal scaling functions in terms of the variable $m^{1/(1+\gamma^\star)} L$ \cite{DelDebbio:2010jy, DelDebbio:2010ze,DeGrand:2011cu}.  Observation of ``curve collapse" in measurements on several volumes can therefore provide a clearer signal that the underlying theory is described by an infrared fixed point, at least over the range of energy scales considered.

It is instructive to compare the situation at $N_f = 10$ to the current state of knowledge for the $N_f = 12$ theory.  Although the spectrum-fitting methods used here cannot distinguish a truly conformal theory from one with a small dynamical breaking scale $\mu \ll 1/L$, they do lead to $\gamma^\star \gtrsim 0.8$, matching expectations for a theory near the edge of the conformal window.  For $N_f = 12$ the same statement applies, except that mass-deformed fits \cite{Fodor:2011tu,Appelquist:2011dp,DeGrand:2011cu} indicate $\gamma^\star \sim 0.4 \ll 1$.  Assuming that $\gamma^\star$ increases monotonically as $N_f$ approaches $N_f^c$ and that the conformal window closes near $\gamma^\star = 1$ \cite{Holdom:1984sk,Yamawaki:1985zg,Appelquist:1986an}, a value of $\gamma^\star \sim 0.4$ indicates that the theory is inside and not particularly close to the edge of the window.  Furthermore, several groups have carried out direct studies of the renormalization-group flow, through running coupling studies or similar techniques \cite{Appelquist:2007hu,Appelquist:2009ty,Hasenfratz:2011xn,Aoki:2012kr,Aoyama:2011ry,Deuzeman:2012pv}.  The observation of backwards renormalization-group flow by some of these groups gives additional evidence that the $N_f = 12$ theory is in the conformal window.

\paragraph*{\textbf{Acknowledgments}}
We thank the LLNL Multiprogrammatic and Institutional Computing program for time on the BlueGene/L supercomputer, along with funding from LDRD 10-ERD-033.  This work has been supported by the U.~S.~Department of Energy under Grant Nos.~DE-FG02-04ER41290 (D.S.), DE-FG02-91ER40676 (R.C.B., M.C., C.R.), DE-FG02-92ER-40704 (T.A.) and Contracts DE-AC52-07NA27344 (LLNL), DE-AC02-06CH11357 (Argonne Leadership Computing Facility), and DE-AC02-07CH11359 (Fermi Research Alliance, LLC), and by the National Science Foundation under Grant Nos.~NSF PHY11-00905 (G.F., M.L., G.V.) and PHY11-25915 (Kavli Institute for Theoretical Physics).

\bibliography{prl-10f}

\appendix

\subsection{Appendix: data tables}

To appear in supplementary journal material.

\begin{center}

\begin{table}[p]
\begin{center}
\begin{tabular}{|c|ccccc|}
\hline
$m_f$&$M_P$&$M_V$&$M_A$&$M_N$&$M_{N^\star}$\\
\hline
0.010&0.1954(44)&0.2442(30)&0.2620(46)&0.3804(59)&0.4277(69)\\
0.015&0.2023(77)&0.2443(40)&0.2836(42)&0.3652(79)&0.4319(18)\\
0.020&0.2252(23)&0.2660(40)&0.3257(23)&0.3933(114)&0.4865(80)\\
0.025&0.2482(19)&0.2855(19)&0.3487(44)&0.4187(55)&0.5075(107)\\
0.030&0.2762(16)&0.3185(24)&0.4064(37)&0.4703(57)&0.6043(115)\\
\hline
\hline
$m_f$&$m_{res}$&$F_P$&$F_V$&$F_A$&$\pbp (\times 10^{-2})$\\
\hline
0.010&0.001737(3)&0.01720(30)&0.03611(71)&0.03479(49)&1.37790(44)\\
0.015&0.001765(4)&0.02289(41)&0.03551(73)&0.03111(80)&1.96148(54)\\
0.020&0.001812(2)&0.02852(63)&0.03728(88)&0.03188(112)&2.54567(85)\\
0.025&0.001855(4)&0.03204(40)&0.03811(48)&0.02838(54)&3.12672(47)\\
0.030&0.001883(4)&0.03772(56)&0.04193(54)&0.03137(69)&3.70510(73)\\
\hline
\end{tabular}
\caption{Ordered-start value for all $N_f = 10$ observables. \label{tab:orddata}}
\end{center}
\end{table}

\begin{table}[]
\begin{center}
\begin{tabular}{|c|ccccc|}
\hline
$m_f$&$M_P$&$M_V$&$M_A$&$M_N$&$M_{N^\star}$\\
\hline
0.010&0.1829(30)&0.2418(51)&0.2686(60)&0.3637(58)&0.4107(88)\\
0.015&0.1889(40)&0.2376(59)&0.3129(34)&0.3630(90)&0.4508(166)\\
0.020&0.2464(28)&0.2902(24)&0.3659(24)&0.4297(48)&0.6118(181)\\
0.025&0.2673(12)&0.3209(20)&0.4190(50)&0.4774(47)&0.8392(706)\\
0.030&0.2844(12)&0.3379(21)&0.4372(64)&0.5053(49)&0.6710(174)\\
\hline
\hline
$m_f$&$m_{res}$&$F_P$&$F_V$&$F_A$&$\pbp (\times 10^{-2})$\\
\hline
0.010&0.001740(4)&0.01981(123)&0.03755(83)&0.03608(139)&1.37796(42)\\
0.015&0.001759(5)&0.02704(45)&0.03611(91)&0.03193(71)&1.96729(61)\\
0.020&0.001790(4)&0.03497(98)&0.04141(59)&0.03341(29)&2.53830(53)\\
0.025&0.001820(3)&0.03794(71)&0.04634(91)&0.03628(122)&3.11588(104)\\
0.030&0.001869(2)&0.03878(71)&0.04502(68)&0.03558(142)&3.69653(82)\\
\hline
\end{tabular}
\caption{Disordered-start value for all $N_f = 10$ observables. \label{tab:orddata}}
\end{center}
\end{table}

\begin{table}[]
\begin{center}
\begin{tabular}{|c|ccccc|}
\hline
$m_f$&$M_P$&$M_V$&$M_A$&$M_N$&$M_{N^\star}$\\
\hline
0.010&0.1891(62)&0.2430(92)&0.2653(150)&0.372(15)&0.419(60)\\
0.015&0.1960(80)&0.2410(95)&0.2982(162)&0.364(16)&0.441(65)\\
0.020&0.2359(66)&0.2781(99)&0.3458(184)&0.411(18)&0.549(79)\\
0.025&0.2578(69)&0.3032(104)&0.3838(208)&0.448(18)&0.673(108)\\
0.030&0.2803(74)&0.3282(113)&0.4218(229)&0.488(19)&0.638(92)\\
\hline
\hline
$m_f$&$m_{res}$&$F_P$&$F_V$&$F_A$&$\pbp (\times 10^{-2})$\\
\hline
0.010&0.001738(10)&0.0185(14)&0.0368(22)&0.0354(27)&1.3779(16)\\
0.015&0.001762(11)&0.0250(15)&0.0358(22)&0.0315(23)&1.9644(21)\\
0.020&0.001801(10)&0.0317(20)&0.0393(23)&0.0326(24)&2.5420(27)\\
0.025&0.001837(11)&0.0350(21)&0.0422(25)&0.0323(25)&3.1213(32)\\
0.030&0.001876(11)&0.0383(23)&0.0435(25)&0.0335(26)&3.7001(38)\\
\hline
\end{tabular}
\caption{Combined values for all $N_f = 10$ observables, using the method described in the text.\label{tab:massdata}}
\end{center}
\end{table}

\end{center}

\end{document}